\documentclass[12pt,reprint,aps,prb,amsmath,amssymb]{revtex4-2}

\usepackage{hyperref}
\usepackage{graphicx}
\usepackage{xcolor}

\hypersetup{colorlinks=true,citecolor=blue}

\begin{document}

\title{Time reversal symmetry protected chaotic fixed point in the quench dynamics of a topological $p$-wave superfluid}

\author{Aidan Zabalo}
\affiliation{Department of Physics and Astronomy, Center for Materials Theory, Rutgers University, Piscataway, NJ 08854 USA}
\author{Emil A. Yuzbashyan}
\affiliation{Department of Physics and Astronomy, Center for Materials Theory, Rutgers University, Piscataway, NJ 08854 USA}

\date{\today}

\begin{abstract}
We study the quench dynamics of a topological $p$-wave superfluid with two
competing order parameters, $\Delta_\pm(t)$. When the system is prepared in the
$p+ip$ ground state and the interaction strength is quenched, only $\Delta_+(t)$
is nonzero. However, we show that fluctuations in the initial conditions result
in the growth of $\Delta_-(t)$ and chaotic oscillations of both order
parameters. We term this behavior phase III'. In addition, there are two other
types of late time dynamics -- phase I where both order parameters decay to zero
and phase II where $\Delta_+(t)$ asymptotes to a nonzero constant while
$\Delta_-(t)$ oscillates near zero. Although the model is nonintegrable, we
are able to map out the exact phase boundaries in parameter space.
Interestingly, we find phase III' is unstable with respect to breaking the time
reversal symmetry of the interaction. When one of the order parameters is
favored in the Hamiltonian, the other one rapidly vanishes and the previously
chaotic phase III' is replaced by the Floquet topological phase III that is seen
in the integrable chiral $p$-wave model.
\end{abstract}

\maketitle

\section{Introduction}
\begin{figure}
	\centering
	\includegraphics{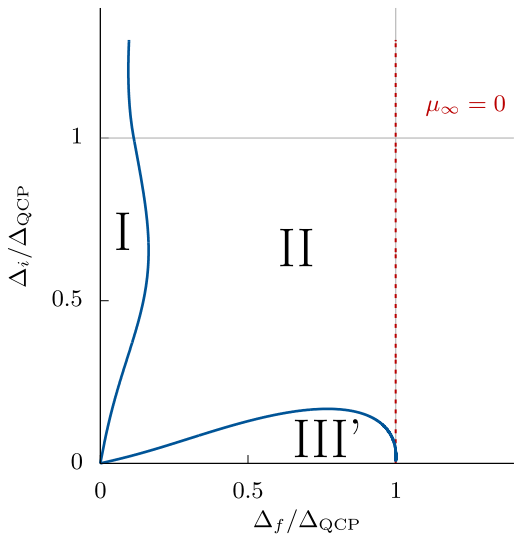}
	\caption{Exact quantum quench phase diagram of a $p$-wave superfluid. Each point represents an interaction quench from a slightly perturbed $p+ip$ ground state, where $\Delta_i$ and $\Delta_f$ are the   ground state $p$-wave order parameter amplitudes for the initial and final couplings.  In equilibrium, there is a quantum critical point at $\Delta=\Delta_\mathrm{QCP}$, which separates the topologically
	nontrivial BCS ($\Delta<\Delta_\mathrm{QCP}$) and topologically trivial BEC ($\Delta>\Delta_\mathrm{QCP}$)    ground states.
	The red dashed line is the nonequilibrium extension of this critical point.
	Away from equilibrium, the system exhibits competing $p\pm ip$
	orders each with its own time dependent complex amplitude $\Delta_\pm(t)$.  In phase I,  both amplitudes decay to zero due to
	dephasing, $\lvert\Delta_{\pm}(t)\rvert\rightarrow 0$. In phase II, the
	$p+ip$ amplitude has damped oscillations and decays to a nonzero constant, while the
	$p-ip$ one shows small oscillations. In phase III', both amplitudes grow and exhibit chaotic
	dynamics. }
	\label{fig:full_PD}
\end{figure}

\begin{figure}
  \centering
  \includegraphics[scale=0.8]{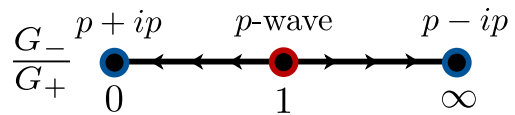}
  \caption{A diagram illustrating the effect of  time reversal symmetry breaking on $p$-wave superfluid dynamics. $G_\pm$ are
  coupling constants of the $p+ ip$ and $p-ip$ interaction terms. The symmetry is broken whenever $G_+\ne G_-$. In much of this
  paper we study a time reversal invariant superfluid where $G_+=G_-$. For a chiral $p$-wave superfluid either $G_-=0$ or $G_+=0$
  leading to a quench phase diagram which differs from Fig.~\ref{fig:full_PD} by a replacement of phase III' with a Floquet topological phase III.
  For $G_+\neq G_-$, we find that the pairing amplitude
  associated with the weaker channel rapidly vanishes. The amplitude of the
  stronger channel survives to late times and its dynamics resemble the quench dynamics of the chiral $p$-wave superfluid.
  Phase III' is thus an unstable fixed point protected by time reversal symmetry.}
  \label{fig:fixedpoint_flow}
\end{figure}
One of the most significant problems challenging our modern understanding of
physics is the characterization of many body systems that are far from
equilibrium. The extent to which conventional tools and frameworks such as the
Landau-Ginzburg-Wilson theory, topological order, and universality remain valid
descriptions of systems out of equilibrium is not readily understood.
Fortunately, in recent years, there has been great progress in the development
of both experimental and theoretical tools that allow us to begin answering such
questions.

Advances in ultra cold atomic systems~\cite{kinoshita2006quantum, lignier2007dynamical, smale2019observation, tang2018thermalization, langen2015experimental, hofferberth2007non, weiler2008spontaneous, widera2008quantum, gring2012relaxation, bloch2008many,regal2004observation, bloch2012quantum, zwierlein2005vortices, zwierlein2004condensation},
quantum devices~\cite{arute2020hartree,gong2021quantum,satzinger2021realizing,cong2021hardwareefficient,monroe2021programmable},
and high frequency pump-probe spectroscopy~\cite{fausti2011light,giannetti2016ultrafast,kampfrath2013resonant,matsunaga2014light,shimano2020higgs,matsunaga2013higgs,demsar2020non} have provided a platform for simulating quantum many body dynamics.
These experiments have shown great promise in their ability to both guide and verify our understanding of
thermalization and nonequilibrium dynamical phases.
There have also been numerical and analytical techniques developed for studying the dynamics of systems
far from equilibrium~\cite{schollwock2005density,rigol2007relaxation,kamenev2010keldysh,polkovnikov2010phase,essler2016quench, calabrese2016quantum,PhysRevLett.96.136801,vasseur2016nonequilibrium,eisert2015quantum}
as well as efforts at defining a notion of nonequilibrium topology~\cite{Foster_2013,foster2014quench,d2015dynamical,Moessner_2017, Yang_2018, Gong_2018, Sun_2018, McGinley_2019, tonielli2020topological, tarnowski2019measuring, Kitagawa_2010, Else_2016, potter_2016, roy_2017, wang_2017}.

In this work, we characterize the late time quantum quench dynamics of a topological 2D
$p$-wave superfluid with two competing order
parameters~\cite{read2000paired,volovik2003universe}. Such a system can, in principle,
be realized in the context of cold atomic gases where an attractive interaction
between identical fermions can be tuned through a Feshbach
resonance~\cite{gurarie2007resonantly,chin2010feshbach,bloch2008many}.
The system is expected to have $p+ip$ and $p-ip$ ground states, where $p\pm ip$ refers to the symmetry of the
superfluid order parameter: $\Delta_\mathbf{p} = (p^x \mp ip^y)\Delta_\pm$,  $p^x$ and $p^y$ are the $x$ and $y$ components
of the 2D momentum $\mathbf{p}$, and
$\Delta_+$ and $\Delta_-$ are the $p+ip$ and $p-ip$ pairing amplitudes, respectively.
In the $p+ip$ ground state $\Delta_-=0$ while in the $p-ip$ ground state $\Delta_+=0$.
Remarkably, the ground state can be tuned across a quantum phase transition by varying the chemical potential, $\mu$, of the system. For $\mu < 0$, the system is in the topologically trivial strong pairing BEC phase while for $\mu > 0$ it is in the topologically nontrivial weak pairing BCS phase.
The transition occurs at the quantum critical point $\mu = 0$ where the ground state pairing amplitude takes on the value $\Delta_\mathrm{QCP}$, see Sec.~\ref{sec:eq_top} for details.

Unfortunately, attempts to experimentally realize such a gas have proven difficult due to the short lifetimes
before losses induced by three-body processes destabilize the
gas~\cite{chin2010feshbach}. However, the ability to tune interactions via
resonances has led to the consideration of using out of equilibrium dynamics
as a means to induce metastable phases. In particular, it was argued that
from a weakly paired $p+ip$ or, equivalently, $p-ip$ ground state it is possible to induce a
Floquet topological superfluid by a sudden interaction quench in the time before the instability
occurs~\cite{foster2014quench}.

Though this seems  promising, a deeper understanding of
the nonequilibrium $p$-wave superfluid is necessary before  conclusions can be drawn.
The degree of fine tuning  required to realize this
behavior has not been understood and, naturally, the question arises as
to whether these dynamics are stable against small fluctuations around the
$p+ip$ ground state due to, e.g., additional interactions, finite temperature, or coupling to the environment. In other words, do deviations from the $p+ip$ ground state
affect the existence of this phase? Instabilities in oscillatory dynamical
phases have been shown to occur in similar models of
superfluids~\cite{Yuzbashyan_2009,Dzero_2009}. There, the instabilities are
driven by spatial, thermal or quantum fluctuations.

Through the use of analytical techniques and numerical simulations we are
able to map out the entire interaction quench phase diagram of the $p$-wave superfluid, see Fig.~\ref{fig:full_PD}.
We start in a slightly perturbed $p+ip$ ground state with initial superfluid interaction strength $G_i$ and then abruptly change the
interaction  to $G_f$. The ground state value of the pairing amplitude $\Delta_{0,+}(G)$ is a monotonic function of the interaction strength $G$ and
we find it convenient to represent the $G_i\rightarrow G_f$ quench as a point with coordinates $(\Delta_i, \Delta_f)$ in the phase diagram,
where $\Delta_i=\Delta_{0,+}(G_i)$ and  $\Delta_f=\Delta_{0,+}(G_f)$. By symmetry, the phase diagram for quenches from a slightly perturbed $p-ip$ ground
state is obtained via a simple interchange of $\Delta_+$ and $\Delta_-$.

Unfortunately, we find that small fluctuations completely destroy the Floquet
topological superfluid. The quench phase diagram of the $p$-wave superfluid
consists of three nonequilibrium phases classified according to the late time
dynamics of the two order parameter pairing amplitudes $\Delta_+(t)$ and
$\Delta_-(t)$. In phase I, both
amplitudes decay to zero due to dephasing. In phase II, $\Delta_{+}(t)$ has
damped oscillations and decays to a nonzero constant while $\Delta_{-}(t)$
stays small and shows persistent nonperiodic oscillations. In phase III', the two order
parameter amplitudes grow and exhibit chaotic oscillations. Surprisingly, even though the $p$-wave Hamiltonian is
nonintegrable, we find an analytic description of the phase boundaries that
is consistent with our numerical simulations. The resulting quantum quench phase diagram is shown in Fig.~\ref{fig:full_PD}. The
quench phase diagram of the chiral $p$-wave model studied in
Ref.~\onlinecite{Foster_2013} is  the same except  our new chaotic and nontopological phase
III' is  replaced with
the Floquet topological phase III~\footnote{\label{note1} There are differences in the shapes of phase boundaries and $\mu_\infty=0$ line between our
 Fig.~\ref{fig:full_PD} and Ref.~\onlinecite{Foster_2013}  due to different cutoff conventions, see the end of Sec.~\ref{sectbound}.}.

To gain insight into the properties of phase III', we consider the limit $\Delta_i\rightarrow0$, which corresponds to the horizontal axis of the quench phase diagram in Fig.~\ref{fig:full_PD}. In this case, the initial state is close to the ground state of a free Fermi gas (the normal state) and
we can understand the short time pairing dynamics by performing a linear stability analysis around this state.
 We find that  both amplitudes grow as $\Delta_\pm(t) \propto e^{\gamma t}$ with the same
rate $\gamma$. For small interaction strengths, $\gamma = \Delta_f\sqrt{2\epsilon_F}$, where $\epsilon_F$ is the   Fermi energy.
As we increase $\Delta_f$ moving towards the phase III'--II transition point along the $\Delta_i=0$ line, $\gamma$ decreases until it vanishes at the transition  becoming purely imaginary afterwards.  We find that the transition  point is at $\Delta_f=\Delta_\mathrm{QCP}$. Thus, three transitions occur at the same value of the ground state pairing amplitude  $\Delta_f=\Delta_\mathrm{QCP}$: (i) the equilibrium BCS--BEC quantum phase transition, (ii) the transition between nonequilibrium phases III' and II for $\Delta_i=0$ and (iii) change in the   stability of the normal state with respect to   superfluid interactions. A rather remarkable byproduct of this analysis is that the equilibrium BCS--BEC quantum phase transition can be defined solely in terms of the stability of the normal state -- the BCS phase is when the normal state is dynamically unstable and the BEC phase is when it is stable.

Finally, we study the effects of time reversal symmetry breaking on the late
time dynamics of the quenched $p$-wave superfluid.
The $p$-wave interactions can be divided into $p+ip$ and $p-ip$ interaction channels.
Time reversal invariance requires that interaction strengths of the two channels be equal to each other, $G_+=G_-=G$.
This is the model we consider throughout this paper (except Sec.~\ref{sec:TRB}) and whose quench phase diagram appears in Fig.~\ref{fig:full_PD}.
The chiral $p$-wave model has either $G_+=0$ or $G_-=0$ and accordingly there is only one nonzero pairing amplitude, $\Delta_+$ or $\Delta_-$.
To better understand the effect of time reversal symmetry breaking, we consider the situation when both couplings are nonzero and unequal, $G_+\ne G_-$, in
Sec.~\ref{sec:TRB}. We find that the pairing amplitude associated with the weaker channel rapidly
vanishes while the amplitude of the stronger channel survives to late times regardless of the initial
state, i.e., the stronger channel always wins.
These late time dynamics closely resemble the quench dynamics of a chiral $p$-wave superfluid, though for a modified set of quench parameters.
The $p$-wave phase III' therefore represents an unstable fixed point
protected by time reversal symmetry, see Fig.~\ref{fig:fixedpoint_flow}. As soon as we make $G_+\ne G_-$, the dynamics that previously lead to phase III' take the system into
the Floquet phase III of Ref.~\onlinecite{Foster_2013}. This result
indicates that it is still possible to observe a quench induced Floquet topological
superfluid phase provided the time reversal symmetry of the interaction term
is explicitly broken.

\section{Equilibrium properties of the BCS Hamiltonian\label{sec:model_bg}}
The simplest realistic 2-D $p$-wave BCS Hamiltonian is given by~\cite{gurarie2007resonantly}
\begin{equation}
  \hat{H} = \sum_{\mathbf{k}} \frac{k^{2}}{m} \hat{c}^{\dagger}_{\mathbf{k}} \hat{c}_{\mathbf{k}} - \frac{2G}{m} \sum_{\mathbf{p}, \mathbf{k}, \mathbf{q}} \mathbf{k} \cdot \mathbf{q} \hat{c}^{\dagger}_{\frac{\mathbf{p}}{2} + \mathbf{k}} \hat{c}^{\dagger}_{\frac{\mathbf{p}}{2} - \mathbf{k}} \hat{c}_{\frac{\mathbf{p}}{2} - \mathbf{q}} \hat{c}_{\frac{\mathbf{p}}{2} + \mathbf{q}}
  \label{eq:bcs_ham_elec}
\end{equation}
where the operator $\hat{c}^{\dagger}_{\mathbf{k}}$ ($\hat{c}_{\mathbf{k}}$)
creates (annihilates) a spinless fermion with momentum $\mathbf{k}$ and $G>0$
is the dimensionless BCS coupling. We will focus only on the interaction
terms with $\mathbf{p}=0$, i.e. the reduced BCS model, and neglect the pair
breaking terms, $\mathbf{p} \neq 0$. This approximation is valid away from
equilibrium as long as the characteristic timescale of the dynamics is less
than the time for pair breaking processes to occur. We expect this to be the
case away from the quantum critical point (and its nonequilibrium extension)
where the chemical potential, $\mu~(\mu_\infty)$,
vanishes~\cite{yuzbashyan2015quantum}.

It is convenient to express Eq.~\eqref{eq:bcs_ham_elec} in terms of Anderson
pseudospins defined through the relationships~\cite{anderson1958random}
\begin{align}
  \begin{split}
    \hat{s}_{\mathbf{k}}^{z} &= \frac{1}{2}\left( \hat{c}^{\dagger}_{\mathbf{k}}\hat{c}_{\mathbf{k}}+\hat{c}^{\dagger}_{\mathbf{-k}}\hat{c}_{\mathbf{-k}} -1\right), \\
    \hat{s}_{\mathbf{k}}^{+} &= \hat{c}^{\dagger}_{\mathbf{k}}\hat{c}^{\dagger}_{\mathbf{-k}}, \\
    \hat{s}_{\mathbf{k}}^{-} &= \hat{c}_{\mathbf{-k}}\hat{c}_{\mathbf{k}}.
  \end{split}
\end{align}
With this replacement, the reduced ($\mathbf{p}=0$) Hamiltonian becomes
\begin{equation}
	\hat{H} = \sum_{\mathbf{k}}'k^{2}\hat{s}^{z}_{\mathbf{k}}-2G\sum_{\mathbf{k},\mathbf{q}}' \mathbf{k}\cdot\mathbf{q}\hat{s}_{\mathbf{k}}^{+}\hat{s}_{\mathbf{q}}^{-}
  \label{eq:pwave_ham}
\end{equation}
where, without loss of generality, we have set $m=1$.
The primed sums indicate that the momenta are restricted to the upper half plane
so that $\mathbf{k} = \{k^{x} \in \mathbb{R}, k^{y} \ge 0\}$ and double counting
is avoided.
It is easily verified that the pseudospins satisfy the usual commutation
relations
$\left[\hat{s}_{\mathbf{k}}^{a},\hat{s}_{\mathbf{q}}^{b}\right] = i\delta_{\mathbf{k},\mathbf{q}}\epsilon^{abc}\hat{s}_{\mathbf{k}}^{c}$.

In a mean-field treatment, the Hamiltonian in Eq.~\eqref{eq:pwave_ham} can
be rewritten as
\begin{align}
  \begin{split}
  \hat{H}_{MF} = \sum_{\mathbf{k}}'k^{2}\hat{s}^{z}_{\mathbf{k}}
  &+ \sum_{\mathbf{k}}' k\left(e^{-i\phi_\mathbf{k}}\Delta_+ + e^{+i\phi_\mathbf{k}}\Delta_-\right)\hat{s}_{\mathbf{k}}^{+} \\
  &+ \sum_{\mathbf{k}}' k\left(e^{+i\phi_\mathbf{k}}\Delta^*_+ + e^{-i\phi_\mathbf{k}}\Delta^*_-\right)\hat{s}_{\mathbf{k}}^{-}
  \end{split}
	\label{eq:mf_H}
\end{align}
where
\begin{align}
  \Delta_{\pm} \equiv -G\sum_{\mathbf{k}}ke^{\pm i\phi_{\mathbf{k}}}\langle \hat{s}_{\mathbf{k}}^{-}\rangle
  \label{eq:def_delta}
\end{align}
is the pairing amplitude associated with the $p\pm ip$ superfluid order
parameter, $\phi_\mathbf{k}$ is the polar angle in the $k^x,k^y$ plane, and
the expectation values are taken with respect to the many-body wavefunction
of the system.

The Heisenberg equations of motion for the operators are
\begin{equation}
  \frac{\mathrm{d}\hat{\mathbf{s}}_\mathbf{k}}{\mathrm{d}t} = \hat{\mathbf{s}}_\mathbf{k} \times \mathbf{H}_\mathbf{k},
  \label{eq:blocheq}
\end{equation}
with $\mathbf{H}_\mathbf{k}$ an effective magnetic field given by
\begin{equation}
  \mathbf{H}_\mathbf{k} =
  \begin{bmatrix}
    -k(e^{-i\phi_\mathbf{k}}\Delta_+ + e^{+i\phi_\mathbf{k}}\Delta_-) + c.c. \\
    -ik(e^{-i\phi_\mathbf{k}}\Delta_+ + e^{+i\phi_\mathbf{k}}\Delta_-) + c.c. \\
    -k^2
  \end{bmatrix}.
  \label{eq:spinB}
\end{equation}
This mean-field treatment is generally  exact for pairing models
such as our Hamiltonian, Eq.~\eqref{eq:mf_H}, where interactions are all to
all~\cite{doi:10.1063/1.523493,Roman:2002dh,PhysRevB.71.094505} and should remain valid away from equilibrium~\cite{Faribault_2009,Wu_unpublished} at times smaller than the Ehrenfest time $t_E$. This time is proportional
to $\sqrt{N_s}$, where $N_s$ is the number of spins (equivalently, number of momenta $\mathbf{k}$), except for quenches from the Fermi gas ground state where
$t_E\propto \log N_s$~\cite{Yuzbashyan_2009}, see also Ref.~\onlinecite{kehrein} for similar results in the transverse field Ising model.

Upon taking the expectation value of both sides of Eq.~\eqref{eq:blocheq}, the
equations of motion reduce to Bloch equations $ \dot{\mathbf{s}}_\mathbf{k} =\mathbf{s}_\mathbf{k}\times \mathbf{H}_\mathbf{k}$   for classical spin variables,
$\mathbf{s}_\mathbf{k} = \langle\hat{\mathbf{s}}_\mathbf{k}\rangle$. More
explicitly, we have the classical equations of motion
\begin{align}
  \begin{split}
    \dot{s}_{\mathbf{k}}^{-} = &-ik^{2} s_{\mathbf{k}}^{-} + 2iks_{\mathbf{k}}^{z} \left( e^{-i\phi_{\mathbf{k}}}\Delta_{+} + e^{+i\phi_{\mathbf{k}}}\Delta_{-} \right), \\
    \dot{s}_{\mathbf{k}}^{z} = &-iks_{\mathbf{k}}^{+}\left( e^{-i\phi_{\mathbf{k}}}\Delta_{+} + e^{+i\phi_{\mathbf{k}}}\Delta_{-}\right) \\
    &+iks_{\mathbf{k}}^{-}\left( e^{+i\phi_{\mathbf{k}}}\Delta_{+}^{*} + e^{-i\phi_{\mathbf{k}}}\Delta_{-}^{*}\right).
    \label{eq:eom}
  \end{split}
\end{align}
In equilibrium, the two order parameter amplitudes have a time dependent phase
that winds with frequency $2\mu$,
\begin{equation}
\Delta_\pm(t) = \Delta_{0,\pm}e^{-2i\mu t},
\label{eq:gs_winding}
\end{equation}
where the amplitude $\Delta_{0,\pm}$ is time independent and $\mu$ is the chemical
potential to be determined self-consistently. This phase arises due to the requirement that the expectation
values in Eq.~\eqref{eq:def_delta} are taken between states which differ by two particles.
To eliminate this evolution we move into the rotating frame,
$s_\mathbf{k}^-\rightarrow s_\mathbf{k}^-e^{-2i\mu t}$. In this frame,
 the field which acts on the spins is static and given by
$\mathbf{h}_\mathbf{k} = \mathbf{H}_\mathbf{k}+2\mu\hat{\mathbf{z}}$.
The spin configuration which minimizes the energy can now be found by aligning each spin parallel to its local magnetic field
\begin{equation}
  \begin{split}
    s_{\mathbf{k}0}^{-} = - \frac{k\left( e^{-i\phi_{\mathbf{k}}}\Delta_{0,+} + e^{+i\phi_{\mathbf{k}}}\Delta_{0,-}\right)}{\sqrt{\left( k^{2}-2\mu\right)^{2}+4k^{2}\lvert e^{-i\phi_{\mathbf{k}}}\Delta_{0,+} + e^{+i\phi_{\mathbf{k}}}\Delta_{0,-} \rvert^{2}}}, \\
    s_{\mathbf{k}0}^{z} = - \frac{k^{2}-2\mu}{2\sqrt{\left( k^{2}-2\mu\right)^{2}+4k^{2}\lvert e^{-i\phi_{\mathbf{k}}}\Delta_{0,+} + e^{+i\phi_{\mathbf{k}}}\Delta_{0,-} \rvert^{2}}}.
  \end{split}
  \label{eq:spin_eq}
\end{equation}
Minimizing with respect to $\Delta_{0,\pm}$ one finds that the absolute minimum
corresponds to one of the two order parameter amplitudes being zero:
$\left\{\Delta_{0,+} \neq 0, \Delta_{0,-}=0\right\}$ for the $p + ip$ ground state
or $\left\{\Delta_{0,+} = 0, \Delta_{0,-} \neq 0\right\}$ for the $p - ip$ ground
state~\cite{gurarie2007resonantly}.
This ground state degeneracy appears due to the presence of time reversal symmetry in the Hamiltonian.

Without loss of generality, we choose to work with the $p+ip$ ground state
and set $\Delta_{0,-} = 0$ in Eq.~\eqref{eq:spin_eq}. The ground state pairing
amplitude and chemical potential can then be determined self-consistently
with the help of Eq.~\eqref{eq:def_delta}
\begin{equation}
	\frac{1}{G} = \sum_{\mathbf{k}}\frac{k^{2}}{\sqrt{\left( k^{2}-2\mu\right)^{2}+4k^{2}\lvert \Delta_{0,+} \rvert^{2}}},
  \label{eq:mf_D}
\end{equation}
and by relating the total particle number $N$ to  $\sum_{\bf k} s_{{\bf k}0}^z$
\begin{equation}
  N = \sum_{\mathbf{k}}\left(1-\frac{k^{2}-2\mu}{\sqrt{\left( k^{2}-2\mu\right)^{2}+4k^{2}\lvert \Delta_{0,+} \rvert^{2}}} \right).
  \label{eq:mf_N}
\end{equation}
It is often more convenient to work with the continuum limit of Eq.~\eqref{eq:mf_D} and Eq.~\eqref{eq:mf_N}.
Introducing a high energy cutoff, $\Lambda$, for a system of size $L$ the equations become
\begin{equation}
	\frac{2\pi}{g} = \int_0^{2\Lambda} \mathrm{d}\epsilon \frac{\epsilon}{\sqrt{\left( \epsilon-2\mu\right)^{2}+4\epsilon\lvert \Delta_{0,+} \rvert^{2}}},
  \label{eq:mf_D_cont}
\end{equation}
and
\begin{equation}
	n = \frac{1}{8\pi}\int_0^{2\Lambda} \mathrm{d}\epsilon \left(1-\frac{\epsilon-2\mu}{\sqrt{\left( \epsilon-2\mu\right)^{2}+4\epsilon\lvert \Delta_{0,+} \rvert^{2}}} \right),
  \label{eq:mf_N_cont}
\end{equation}
where we have performed the integral over the polar angle, $\phi_\mathbf{k}$,
and defined $\epsilon = k^2$, $g = G L^2/4$, and $n = N/L^2$.
We evaluate these integrals in Appendix~\ref{apdx:integrals}.

\subsection{Equilibrium topology\label{sec:eq_top}}
An important feature of the $p$-wave superfluid ground state is that it can
be tuned across a topological phase transition by varying the chemical
potential. In the weak pairing BCS phase ($\mu > 0$), the system is
topologically nontrivial and can support chiral Majorana edge modes while in
the strong pairing BEC phase ($\mu<0$), the system is topologically
trivial~\cite{read2000paired,alicea2012new,gurarie2007resonantly}. At the
quantum critical point separating the two phases ($\mu = 0$) the
quasiparticle spectrum becomes gapless. The corresponding value of the order
parameter amplitude at the critical point, $\Delta_\mathrm{QCP}$, can be
determined from Eq.~\eqref{eq:mf_N_cont} to give
 \begin{equation}
\Delta_\mathrm{QCP} = \sqrt{\frac{-4\pi n}{\mathcal{W}_{-1}\left[-\frac{2e\pi
n}{\Lambda}\right]}},
\end{equation}
where $\mathcal{W}_{-1}$ is the $k = -1$
branch of the Lambert $\mbox{W-function}$.

To see how this transition comes about, we
can look at the topological invariant characterizing the two phases.
There are two possible formulations of the invariant based on the winding of the two vector fields underlying the problem.
One definition of the invariant can be given in terms of the winding of the
static magnetic field, $\mathbf{h}_\mathbf{k}$, which acts on the pseudospins.
The field winding number is defined as
\begin{equation}
  W = \frac{1}{8\pi}\epsilon_{ab}\int \mathrm{d}^2\mathbf{k}~
  \mathbf{\hat{h}}_\mathbf{k} \cdot
  \left(\partial_{a} \mathbf{\hat{h}}_\mathbf{k} \times
  \partial_{b} \mathbf{\hat{h}}_\mathbf{k}\right),
  \label{eq:field_winding_def}
\end{equation}
with $\mathbf{\hat{h}}_\mathbf{k} = \frac{\mathbf{h}_\mathbf{k}}{\lvert\mathbf{h}_\mathbf{k} \rvert}$.
In both the BCS and BEC ground states $\mathbf{h}_\mathbf{k}$ is given by
\begin{equation}
\begin{split}
\mathbf{h}_\mathbf{k} = \mathbf{H}_\mathbf{k}+2\mu\hat{\mathbf{z}}=-(2k\cos\phi_{\bf k}\Delta_{0,+})\hat{\mathbf{x}}-\\
 (2k\sin\phi_{\bf k}\Delta_{0,+})\hat{\mathbf{y}}- (k^2-2\mu)\hat{\mathbf{z}},
\end{split}
\label{grsth}
\end{equation}
where, without loss of generality, we have taken $\Delta_{0,+}$ to be real.

Computing the integral yields
\begin{equation}
  W = \frac{1}{2} \left[ 1 + \mathrm{sgn}(\mu)\right].
  \label{eq:field_winding}
\end{equation}
For $\mu > 0$, the field winding number gives $W = 1$ while for $\mu < 0$
it gives $W = 0$. At the critical point, the field winding number is not
a well defined quantity.

An alternative definition of the topological invariant can be given in terms of the winding of the pseudospins, $\mathbf{s}_\mathbf{k}$.
The pseudospin winding number is defined as
\begin{equation}
  Q \equiv \frac{1}{8\pi}\epsilon_{ab}\int \mathrm{d}^2\mathbf{k}
  \frac{\mathbf{s}_\mathbf{k} \cdot
  \left(\partial_{a} \mathbf{s}_\mathbf{k} \times
  \partial_{b} \mathbf{s}_\mathbf{k}\right)}
  {\lvert \mathbf{s}_\mathbf{k}\rvert^3}.
\end{equation}
Substituting the ground state configuration from Eq.~\eqref{eq:spin_eq} with
$\Delta_{0,-} = 0$ and computing the integral gives  $Q=W$
since in the ground state each spin is parallel to its local field.

The equivalence between these two definitions of the topological invariant
only holds in equilibrium. The pseudospin winding number, $Q$, depends only
on the initial state of the system and is conserved throughout the dynamics
so it is of little interest to us~\cite{Foster_2013}. On the other hand, the
field winding number, $W,$ is encoded in the asymptotic dynamics of the system and can
change after a quench across the quantum critical point.
For states in which $\Delta_\pm(t)$ has the form in Eq.~\eqref{eq:gs_winding},
our   $W$ coincides with   the topological invariant
introduced in Ref.~\onlinecite{volovik2003universe} in terms of the retarded single particle
Green's function~\cite{Foster_2013}. Therefore, it can be argued that for such states,
it is $W$ (not $Q$) that determines the presence of Majorana edge modes in the post quench
asymptotic state~\cite{Foster_2013,foster2014quench}.

 \section{Quench dynamics}
In this work, we are interested in studying the dynamics of the two superfluid order
parameters after a sudden quench of the interaction strength. We consider a
system that is initially prepared in a state arbitrarily close to the $p+ip$ ground
state of the mean-field $p$-wave Hamiltonian Eq.~\eqref{eq:mf_H} for some initial coupling $G_{i}$.
The interaction strength is then instantaneously changed to $G_f$ and the system
evolves as a superposition of eigenstates of the new Hamiltonian.
Below we explore the quench phase diagram for various values of
$G_i$ and $G_f$. We parameterize a quench through the use of quench
coordinates of the form $\left\{ \Delta_i,\Delta_f\right\}$, where
$\Delta_{i,f}$ denotes the value of $\Delta_{0,+}$ in the pure $p+ip$ ground
state of the Hamiltonian with interaction strength $G_{i,f}$.

\subsection{Pure $p+ip$ dynamics \label{sec:chiral_dynam}}
\begin{figure*}[hbt]
	\centering
    \includegraphics{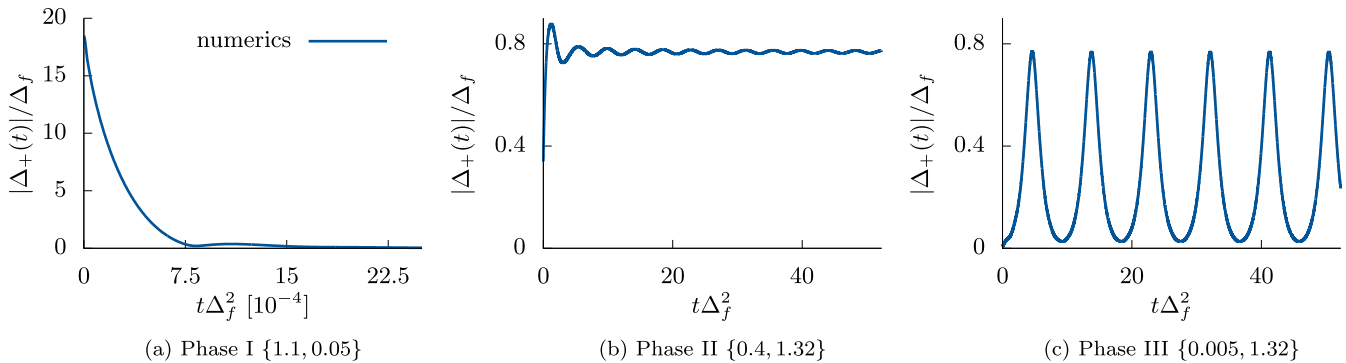}
	\caption{The magnitude of the $p+ip$ order parameter amplitude,
	$\lvert\Delta_+(t)\rvert$, for various interaction quenches, $G_i\to G_f$, in the  $p+ip$
	model Eq.~\eqref{eq:chiral_ham}. The behavior for the full $p$-wave Hamiltonian is the same when
	quenched exactly from the $p+ip$ ground state. The quench coordinates
	  $\left\{\Delta_i, \Delta_f\right\}$ correspond to the ground
	state value of the order parameter amplitude, $\Delta_{0,+}$, for the initial and final
	 couplings $G_i$ and $G_f$, respectively. In phase I, the
	order parameter amplitude decays to zero due to dephasing, in phase II it
	exhibits power law damped oscillations and decays to a constant, and in
	phase III it exhibits persistent oscillations. The numerical simulations were performed on Eq.~\eqref{eq:eom_chiral} for $N=50,000$ pseudospins, see Appendix~\ref{apdx:num_sim} for details.
	 }
	\label{fig:chiral_quenches}
\end{figure*}
We begin by considering the dynamics of the two order parameters following a
quench from the exact $p+ip$ ground state -- Eq.~\eqref{eq:spin_eq} with $\Delta_{0,-} = 0$.
For such an initial state, the dynamics significantly simplify and
it can formally be shown that they are equivalent to those generated from the pure $p+ip$
Hamiltonian~\cite{Foster_2013}. To see this, we note that any time dependent $p+ip$ state
can be written in the form
\begin{equation}
  s_{\mathbf{k}}^- \equiv e^{-i\phi_\mathbf{k}}s_k^-, \quad s_{\mathbf{k}} \equiv s_k^z.
  \label{eq:p_ip_state}
\end{equation}
Substituting Eq.~\eqref{eq:p_ip_state} into Eq.~\eqref{eq:def_delta}, we find
\begin{align}
  \begin{split}
    \Delta_- &= -G\sum_\mathbf{k} ke^{-2i\phi_\mathbf{k}}s_k^- \\
    &= -\frac{g}{2\pi^2}\int_0^{2\Lambda}\epsilon s^-(\epsilon)\mathrm{d}\epsilon
    \int_0^\pi e^{-2i\phi} \mathrm{d}\phi = 0,
  \end{split}
  \label{identity}
\end{align}
where in the second line we have taken the continuum limit and performed the
integral over $\phi$. The above equation shows that for any $p+ip$ initial
state $\Delta_-$ remains zero throughout the entire evolution.

A similar substitution into the equations of motion Eq.~\eqref{eq:eom} gives
\begin{align}
  \begin{split}
    \dot{s}_{k}^{-} = &-ik^{2} s_{k}^{-} + 2iks_{k}^{z}\Delta_{+}, \\
    \dot{s}_{k}^{z} = &-ik\left(s_{k}^{+}\Delta_{+} - s_{k}^{-}\Delta_{+}^{*}\right).
    \label{eq:eom_chiral}
  \end{split}
\end{align}
Eq.~\eqref{eq:eom_chiral} is identical to the equations of motion that are
generated by the $p+ip$ Hamiltonian. The $p+ip$ Hamiltonian is a truncated
version of the full $p$-wave Hamiltonian Eq.~\eqref{eq:bcs_ham_elec} where by writing,
\begin{equation}
  \mathbf{k}\cdot\mathbf{q} = \frac{1}{2}\left[(k^{x}-ik^{y})(q^{x}+iq^{y}) + (k^{x}+ik^{y})(q^{x}-iq^{y})  \right],
  \label{eq:kdotq}
\end{equation}
 discarding the second term, keeping only $\mathbf{p}=0$ terms in the interaction as before, and rewriting the fermionic operators
in terms of the Anderson pseudospins we arrive at
\begin{equation}
	\hat{H} = \sum_{\mathbf{k}}'k^{2}\hat{s}^{z}_{\mathbf{k}}-G\sum_{\mathbf{k},\mathbf{q}}' (k^x-ik^y)(q^x+iq^y)  \hat{s}_{\mathbf{k}}^{+}\hat{s}_{\mathbf{q}}^{-}.
  \label{eq:bcs_chiral_ham_elec}
\end{equation}
The mean field Hamiltonian simplifies to
\begin{equation}
  \hat{H}_{MF}^{p+ip} = \sum_{\mathbf{k}}'k^{2}\hat{s}^{z}_{\mathbf{k}}
  +\sum_{\mathbf{k}}' k\left(e^{-i\phi_\mathbf{k}}\Delta_+\hat{s}_{\mathbf{k}}^{+}
  +e^{+i\phi_\mathbf{k}}\Delta_+^*\hat{s}_{\mathbf{k}}^{-}\right).
  \label{eq:chiral_ham}
\end{equation}
The equations of motion generated by this Hamiltonian after applying Eq.~\eqref{eq:p_ip_state} are  Eq.~\eqref{eq:eom_chiral}.  Note that we can alternatively start from a $p-ip$ ground state and follow a
similar logic. The result is  again
Eq.~\eqref{eq:eom_chiral}, but with the replacement
$\Delta_+\rightarrow\Delta_-$. The $p-ip$ Hamiltonian analogous to
Eq.~\eqref{eq:bcs_chiral_ham_elec} would then correspond to discarding the
first term in Eq.~\eqref{eq:kdotq}.

This simplification is crucial in understanding the $p+ip$ dynamics of the
$p$-wave superfluid as the mean field $p+ip$ model was shown to be classically
integrable using a Lax vector construction~\cite{Foster_2013}, which implies Lax pair representation of the equations of motion Eq.~\eqref{eq:eom_chiral},
see Ref.~\cite{rylands2021loschmidt} for details. By studying the behavior of the isolated
roots of the Lax vector norm, the quench phase diagram can be mapped
out~\cite{yuzbashyan2006relaxation, yuzbashyan2006dynamical, Foster_2013,yuzbashyan2015quantum}. We repeat this procedure in Appendix~\ref{apdx:lax} for the parameters used in our numerical
simulations. The resulting phase diagram contains three distinct dynamical
phases characterized by the late time behavior of the order parameter
amplitude $\Delta_+(t)$.

In phase I, the order parameter amplitude decays to zero due to dephasing,
$\lvert\Delta_{+}(t)\rvert\rightarrow 0$. At late times, the spins precess with
frequencies  $k^{2}$ around the $z$-axis and the system is in a gapless
superconducting state. This is different from the normal
state which would have all spins aligned along the $z$-axis.

In phase II, $|\Delta_{+}(t)|$ exhibits damped oscillations and decays to a
nonzero constant, $\lvert\Delta_{+}(t)\rvert\rightarrow\Delta_{\infty}>0$.
The order parameter amplitude has a time dependent phase that winds with frequency $2\mu_\infty$, i.e.,
\begin{equation}
\Delta_+(t) = \Delta_\infty e^{-2i\mu_\infty t},
\end{equation}
similar to its ground state behavior in Eq.~\eqref{eq:gs_winding}.
This suggests that that $\mu_\infty$ is an out of equilibrium chemical potential whose value is determined
by the details of the quench.
In the rotating frame, the pseudospins precess around an effective field
$$
  \mathbf{h}_\mathbf{k} =-(2k\cos\phi_{\bf k} \Delta_\infty) \hat{\bf x}-(2k\sin\phi_{\bf k} \Delta_\infty) \hat{\bf y}-(k^2 - 2\mu_{\infty})\hat{\bf z}.
$$
 The expression for the field is of the same form as in the ground state, see Eq.~\eqref{grsth}, and therefore,
 \begin{equation}
  W = \frac{1}{2}\left[ 1 + \mathrm{sgn}(\mu_{\infty})\right].
\end{equation}
This is similar to the ground state result, Eq.~\eqref{eq:field_winding}, but
with the replacement $\mu\rightarrow\mu_\infty$.  We conclude that the change
in $W$ at $\mu_\infty = 0$ marks the nonequilibrium extension of the quantum
critical point. In Appendix~\ref{apdx:beta_qcp}, we show that $\mu_\infty = 0$ defines a straight vertical line $\Delta_f=\Delta_\mathrm{QCP}$ in the quench phase diagram \cite{Note1} as shown in Fig.~\ref{fig:full_PD}.

In phase III, the order parameter amplitude undergoes persistent oscillations
that can be described in terms of elliptic functions.
Additionally, it was shown in Ref.~\onlinecite{foster2014quench} that in multiple
regions of this phase there are crossing edge states suggesting that the
entire phase is Floquet topological. The order parameter dynamics for each of
the phases are shown in Fig.~\ref{fig:chiral_quenches} using a numerical
simulation of Eq.~\eqref{eq:eom_chiral}. For details regarding the numerics
see Appendix~\ref{apdx:num_sim}.

It is somewhat surprising that the quench dynamics of the full $p$-wave Hamiltonian can be
solved exactly for quenches from a pure $p+ip$ ground state given that this Hamiltonian is nonintegrable. To what extent
the dynamics and topological structure survive infinitesimally small
perturbations from this fine tuned initial state is studied in the following
sections.

\section{Full p-wave dynamics\label{sec:chiral_perturb}}

Naturally, the system cannot be prepared in a $p+ip$ ground state, where
$\Delta_{0,-} = 0$, with absolute purity. Inevitably, there are fluctuations
around this state that can come from a variety of sources. We
will see below that  pure  $p+ip$ oscillatory dynamics are unstable towards exponential
growth of the competing $p-ip$ order associated with $\Delta_{0,-}$. Therefore, the precise mechanism of the
fluctuations is unimportant as long as they seed a nonzero initial
$\Delta_{0,-}$. To understand the effect of such fluctuations we study quenches
from an initial state described by Eq.~\eqref{eq:spin_eq} with
$0<\Delta_{0,-}\ll\Delta_{0,+}$.
We find that, in the limit $\Delta_{0,-} \to 0$, the phase diagram remains largely
unaffected with the exception of phase III, wherein $\Delta_-(t)$ undergoes an unstable growth and the smooth oscillatory dynamics of $\Delta_+(t)$ are destroyed.

\begin{figure*}
  \centering
  \includegraphics{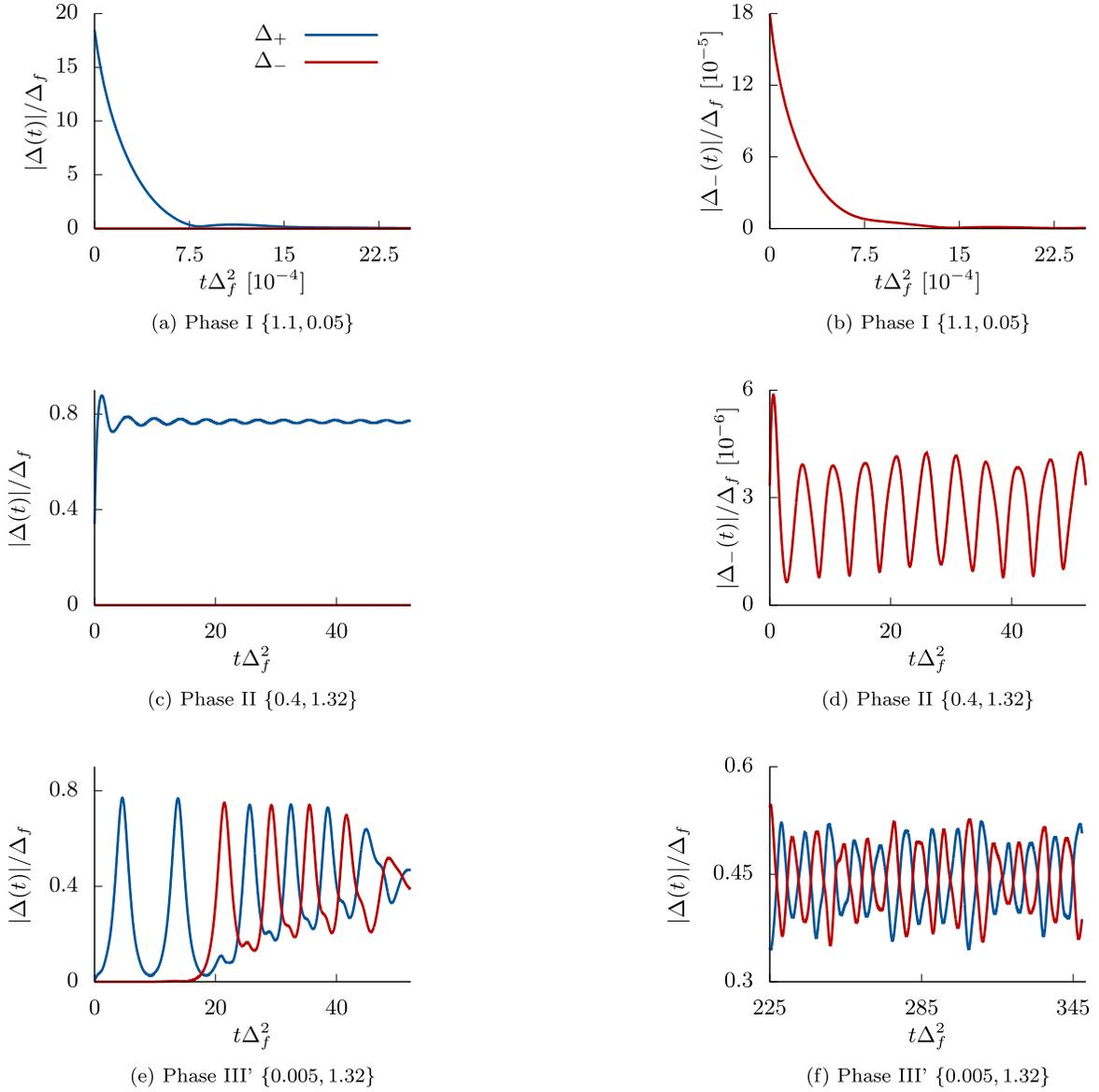}
  \caption{Dynamics of order parameter amplitudes $\lvert \Delta_{+}(t)
  \rvert$ and $\lvert \Delta_{-}(t) \rvert$ for a quenched $p$-wave
  superfluid starting from a slightly perturbed $p+ip$ ground state. In phase
  I, (a) and (b), both order parameter amplitudes decay to zero. In phase II,
  (c) and (d), $\Delta_+(t)$ has damped oscillations and decays to a nonzero
  constant while $\Delta_-(t)$ has small and smooth oscillations. In both of
  these phases, $\Delta_{-}(t)$ is bounded and the dynamics of
  $\Delta_{+}(t)$ are unchanged. In phase III', (e) and (f), the $\Delta_-$
  perturbation grows and destroys the persistent oscillations exhibited by
  the $p+ip$ model, see Fig.~\ref{fig:chiral_quenches}c. At late times, the irregular oscillations of both order
  parameter amplitudes are out of phase with each other.
  As in the previous figure, the numbers in the curly brackets are the ground state values of $\Delta_{0,+}$ for the initial and final couplings. The numerical simulations were performed on Eq.~\eqref{eq:eom} for $N=200,025$ pseudospins with initial conditions Eq.~\eqref{eq:spin_eq} and $\Delta_{0,-}=10^{-5}\Delta_{0,+}$, see Appendix~\ref{apdx:num_sim} for details.}
  \label{fig:full_quenches}
\end{figure*}

\subsection{Stable phases I and II}
For quenches whose coordinates lie within phases I or II of the $p+ip$ phase
diagram, the dynamics of the full $p$-wave Hamiltonian closely mirror that of
the truncated $p+ip$ model, as shown in Fig.~\ref{fig:full_quenches}. In
phase I, we find that both $\Delta_{+}(t)$ and $\Delta_{-}(t)$ decay to zero
in a similar fashion. As before, at late times the pseudospins precess around
the the $z$-axis with frequencies $k^{2}$. In phase II, $\Delta_{+}(t)$
exhibits damped oscillations and decays to a constant. On the other hand,
$\Delta_{-}(t)$ exhibits smooth oscillations that do not seem to decay. The
perturbation to the initial state does not affect the dynamics significantly
since at $t=0$ we have $\Delta_{-}(0)/\Delta_{+}(0) \ll 1$ and as
$t\rightarrow\infty$ the size of $\Delta_{-}(t)$ remains bounded and is
negligible as compared to $\Delta_{+}(t)$. The asymptotic state of the system is
largely unaffected by the perturbation to the initial state and thus we
expect the nonequilibrium topology of phases I and II to remain unchanged.

\subsection{Unstable phase III}

From numerical simulations we find the persistent oscillations observed in phase
III are {\it{not}} stable to the $\Delta_-$ perturbation. Unlike in phases I
and II, the $p-ip$ order parameter amplitude exhibits unstable growth,
and after some delay time,
$\tau$, the persistent oscillations are destroyed and the system enters a regime
of chaotic dynamics which we label as phase III'.
Since quenches from $g_i=0$ to not too large $g_f$ belong to phase III', we can understand the behavior in this phase by linearizing the equations of
motion about the unpaired (free Fermi gas) ground state, which has $\Delta_{i} = 0$ and $\mu_i = 2\pi n$. In the continuum limit,
the growth exponent $\gamma$ is defined through the self-consistency equation of
the linearized problem
\begin{equation}
	\frac{2\pi}{g_f} = \int_0^{2\Lambda} \mathrm{d}\epsilon \frac{\epsilon\,\mathrm{sgn}(\epsilon-2\mu_i)}{\epsilon - \zeta}
\end{equation}
where $\zeta = \omega + i\gamma$.  We can rewrite $g_{f}$ in terms of
$\Delta_{f}$ through Eq.~\eqref{eq:mf_D_cont} and perform the corresponding
integrals on both sides of the equation.  Discarding terms of order $O(\Lambda^{-1})$, we obtain
\begin{equation}
    \zeta \log\left[\frac{-(2\mu_i-\zeta)^2}{2\Lambda\zeta}\right] =
    2\mu_f\log\left[\frac{\Delta_f^2 + 2\lvert\mu_{f}\rvert\Theta(-\mu_{f})}{2\Lambda}\right].
    \label{eq:lin_zeta}
\end{equation}
To simplify the analysis we may consider a quench to weak final pairing
$\Delta_{f} \ll \Delta_\mathrm{QCP}$.  In this case we can take
$\mu_{i}\sim\mu_{f}\gg\Delta_{f}^{2}$ and look for solutions of the form
$\zeta = 2\mu_{i} + i\gamma$ for which we find
$\gamma = \Delta_{f}\sqrt{2\mu_{i}}$. From numerical simulations we find the
initial growth of $\Delta_{-}(t)$ to be suppressed by nonlinear effects and
occurring at later times than predicted by the linear analysis.

We can also use Eq.~\eqref{eq:lin_zeta} to determine the  point
separating nonequilibrium phases II and III' along the $\Delta_{f}$ axis. As we cross from
phase III' to II, the imaginary part of $\zeta$ vanishes, removing the
exponential growth of $\Delta_{-}(t)$. Setting $\gamma = 0$, we see that the
only possible solutions are for $\omega \le 0$. Rewriting
Eq.~\eqref{eq:lin_zeta} in terms of $\omega$ we have
\begin{equation}
  -\lvert\omega\rvert\log\left[\frac{(2\mu_i+\lvert\omega\rvert)^2}{2\Lambda\lvert\omega\rvert}\right] =
  2\mu_f\log\left[\frac{\Delta_f^2 + 2\lvert\mu_{f}\rvert\Theta(-\mu_{f})}{2\Lambda}\right].
\end{equation}
In order to have real solutions for $\omega$, we must require that $\mu_f \le
0$. This first happens when $\Delta_{f} = \Delta_\mathrm{QCP}$ where $\mu_f =
0$ and, therefore, the quantum critical point defines the separation point, see
Fig.~\ref{fig:full_PD}.

Interestingly, this stability analysis also has important implications for the equilibrium physics of the $p$-wave superfluid.  Indeed,  we
investigated here the stability of the unpaired ground  state (normal state) with respect to the $p$-wave Hamiltonian Eq.~\eqref{eq:pwave_ham} with coupling $G=G_f$ and found that this state is unstable in the BCS phase, when the ground state pairing amplitude is smaller then $\Delta_\mathrm{QCP}$, and stable in the BEC phase, when it exceeds $\Delta_\mathrm{QCP}$. Thus, the change in stability of the normal state state identifies the  BCS-BEC quantum phase transition. We note also that our analysis of the free energy shows that even though the normal state
is dynamically stable in the BEC state, it is not a local minimum of the free energy.

\subsection{Signatures of chaos}
\begin{figure}
    \includegraphics{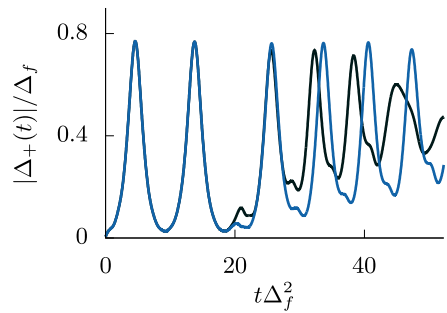}
    \caption{For a quench in phase III', two trajectories whose initial conditions
    vary only slightly will rapidly diverge. The $\Delta_-$ perturbations are given
    by $\Delta_{0,-} = 1\cdot 10^{-5}\Delta_{0,+}$ and
    $\Delta_{0,-} = 2\cdot 10^{-5}\Delta_{0,+}$ for a quench with coordinates
    $\{\Delta_i,\Delta_f\}=\{0.005, 1.32\}$.
    Other parameters and conventions are the same as in the previous figure.}
    \label{fig:vary_ic}
\end{figure}

\begin{figure*}
    \includegraphics{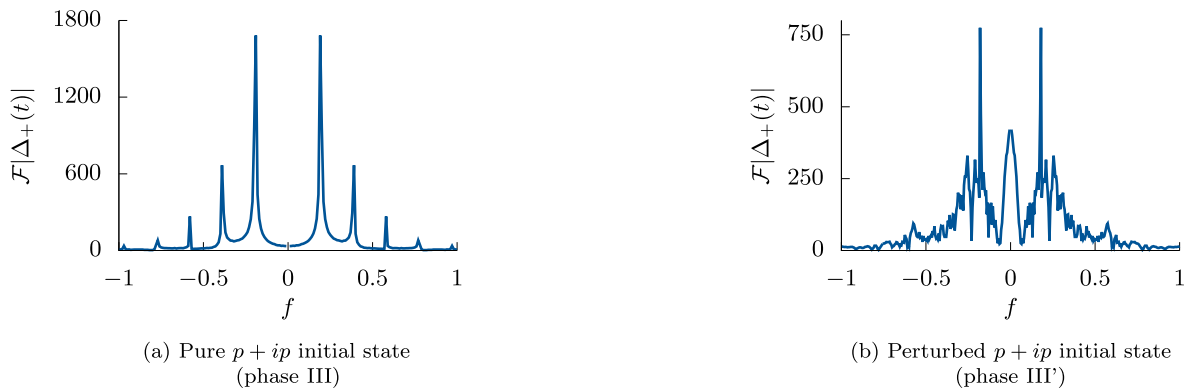}
  \caption{Fourier transform  of the pairing amplitude $\lvert\Delta_{+}(t)\rvert$ for a quench
      from (a)  pure $p+ip$ ground state
    and (b) perturbed $p+ip$ ground state in the full $p$-wave model. The perturbation and quench parameters are the same as in Fig.~\ref{fig:vary_ic}. Spectrum (a) shows a single independent frequency characteristic of the Floquet topological superfluid phase III of the truncated $p+ip$ model. The perturbation  makes the spectrum continuous in (b) suggesting chaotic dynamics in phase III', which replaces phase III in the full $p$-wave model, and indicating the destruction of the
 Floquet topological order.}
	\label{fig:fourier_comp}
\end{figure*}

The late time dynamics in phase III' appear to be chaotic in contrast to
phase III of the  $p+ip$ model. We believe this is because the full
$p$-wave model, Eq.~\eqref{eq:pwave_ham}, is nonintegrable unlike its
truncated $p+ip$ counterpart, Eq.~\eqref{eq:bcs_chiral_ham_elec}. One piece
of evidence of chaos is that the dynamics are sensitive to the initial
conditions, as demonstrated in Fig.~\ref{fig:vary_ic}, where varying the
magnitude of the $\Delta_-$ perturbation yields vastly different late time
trajectories. This behavior makes it difficult to obtain reliable numerical
results at late times since increasing the number of spins effectively modifies the initial conditions.
However, the qualitative behavior, i.e., the appearance of irregular oscillations, is the same.
Additionally, the Fourier spectrum of the time series changes
from a discrete frequency in the classically integrable $p+ip$ case to a
continuous spectrum in the full $p$-wave case as shown in
Fig.~\ref{fig:fourier_comp}.

\subsection{Phase III' topology}
At late times, the irregular oscillations of the two order parameters are out
of phase with one another. The system dynamics are no longer periodic and the
Floquet topological superfluid phase seen in Ref.~\cite{foster2014quench} is therefore
destroyed. Due to the chaotic nature of the dynamics there  is no remnant of
topological order in any known sense in the system.

\subsection{Determination of the phase boundaries\label{sectbound}}
Although the full $p$-wave Hamiltonian is nonintegrable, it is possible to
determine the phase boundaries analytically. We have seen that the dynamics
of $\Delta_+(t)$ in phases I and II are nearly identical to those of the
$p+ip$ model. We can therefore expect this phase boundary to remain
unchanged. In phase III', we have seen that there is an instability of
$\Delta_-(t)$ that leads to chaotic dynamics of both the orders at late
times. However, this instability is confined to the phase III region of the
$p+ip$ model and again there is no choice but for the phase
boundary to remain unchanged between phases II and III'. The exact phase
boundaries for the full $p$-wave model can then be determined by performing
the same analysis of the Lax roots outlined in Ref.~\onlinecite{Foster_2013}. This
is done in Appendix~\ref{apdx:lax} and the result is used to generate the phase
diagram shown in Fig.~\ref{fig:full_PD}. These phase boundaries are in
agreement with direct numerical simulations of the dynamics. We also expect the
$\mu_\infty = 0$ line to remain unchanged since it lies entirely within the stable phase II.

There are differences between shapes of the various lines in our
phase diagram as compared to that of Ref.~\onlinecite{Foster_2013} due to the  cutoff prescription of Ref.~\onlinecite{Foster_2013}, which incorporates the chemical potential, i.e., replaces our $\Lambda$ with
$\Lambda + \mu$ with $\mu$ being the equilibrium chemical potential. Because $\mu$ depends on the coupling constant, this cutoff changes as a result of a quench. The cutoff represents an energy scale governed by physics at higher energies, such as, e.g., the Debye energy in phonon mediated superconductors. It is  more natural to keep this energy scale fixed and unaffected
by the quench. We therefore use
   a fixed  cutoff  throughout the quench phase diagram, which also results in simpler and more intuitive line shapes. In particular, our phase III' boundary  terminates at
$\Delta_\mathrm{QCP}$ along the $\Delta_f$ axis and our $\mu_\infty = 0$ line
is a straight   vertical line defined by the equation $\Delta_f=\Delta_\mathrm{QCP}$.

\section{Stability of the p-wave superfluid with respect to time reversal symmetry breaking~\label{sec:TRB}}
\begin{figure*}
  \centering
  \includegraphics{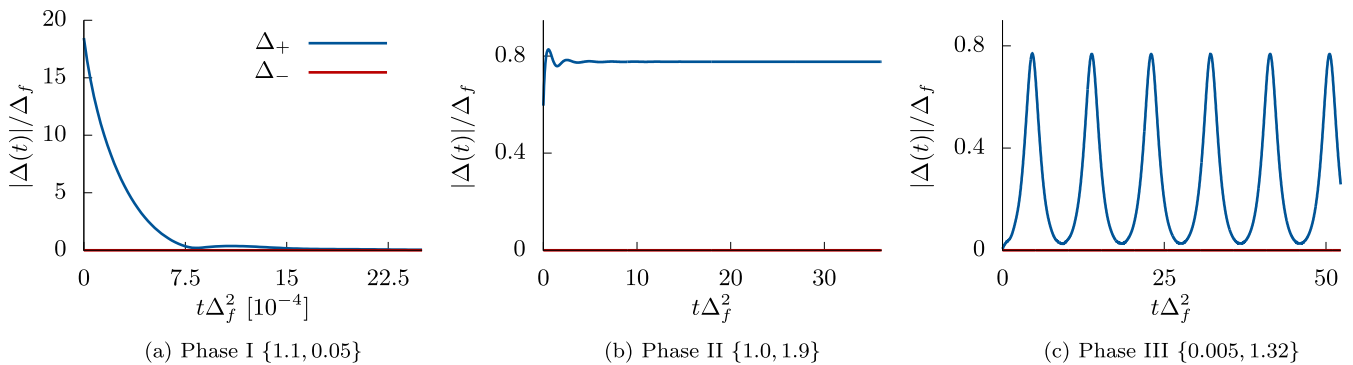}
  \caption{The time reversal symmetry of the $p$-wave interaction is broken
  by making one of the interaction channels stronger, here we have used $G_+
  > G_-$. The figures shown are representative of the two order parameter
  amplitudes in all three phases for a quench from a slightly perturbed
  $p+ip$ ground state. The $\Delta_-$ perturbation is irrelevant to the late
  time dynamics of $\Delta_{+}(t)$ as it rapidly vanishes in all three
  phases. The resulting dynamics of $\Delta_{+}(t)$ closely resemble the pure $p+ip$ quench dynamics. Notations and unspecified parameters are the same as in Fig.~\ref{fig:full_quenches}.}
  \label{fig:dom_right}
\end{figure*}

\begin{figure*}
  \centering
  \includegraphics{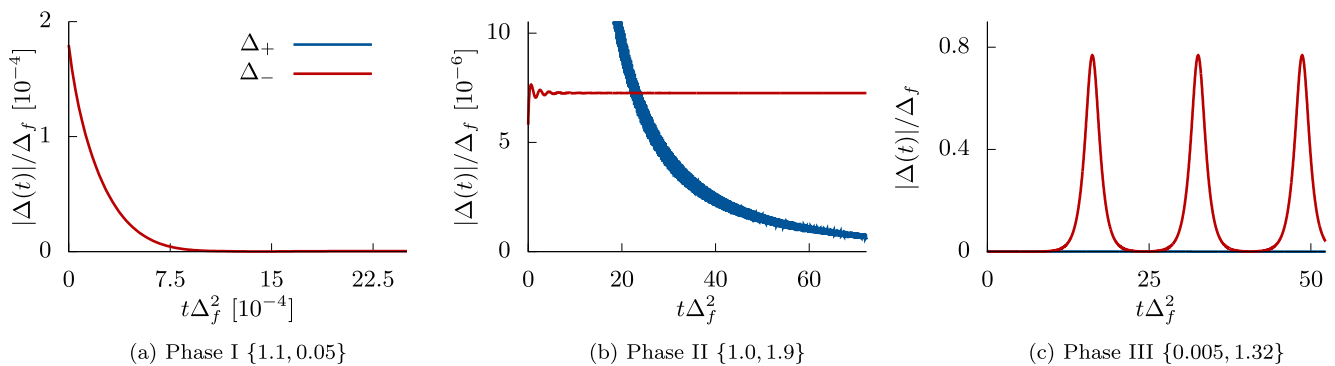}
  \caption{Same as Fig.~\ref{fig:dom_right} but with $G_{+} < G_{-}$. Even
  though the $\Delta_-$ perturbation starts at $10^{-5}\Delta_{0,+}$, it is the
  one whose dynamics survive out to late times. The $p+ip$ order parameter
  amplitude vanishes in all three phases and the dynamics of $\Delta_{-}(t)$
  closely resemble the pure $p-ip$ quench dynamics, though for a different set of quench coordinates.
  Note, in (a), $\Delta_{+}(t)$ vanishes
  slower than $\Delta_{-}(t)$ due to the relative sizes of $G_+$ and $G_-$
  and so it does not show up with the scales used in the figure.}
  \label{fig:dom_wrong}
\end{figure*}

In the previous section, we have seen that the phase III dynamics are unstable
to a perturbation of the pure $p+ip$ initial state. To understand the nature of this
instability we study a variation of the Hamiltonian in Eq.~\eqref{eq:pwave_ham} which
explicitly breaks time reversal symmetry due to an asymmetry in the coupling.
To each interaction channel we associate a distinct coupling constant, $G_+\neq G_-$,
\begin{equation}
\begin{split}
	\hat{H}^\mathrm{asym} = \sum_{\mathbf{k}}'k^{2}\hat{s}^{z}_{\mathbf{k}}-G_+\sum_{\mathbf{k},\mathbf{q}}' (k^x-ik^y)(q^x+iq^y)  \hat{s}_{\mathbf{k}}^{+}\hat{s}_{\mathbf{q}}^{-} \\
	-G_-\sum_{\mathbf{k},\mathbf{q}}' (k^x+ik^y)(q^x-iq^y)  \hat{s}_{\mathbf{k}}^{+}\hat{s}_{\mathbf{q}}^{-}
\end{split}
  \label{eq:bcs_asym_ham}
\end{equation}
and define new order parameter amplitudes
 \begin{equation}
  \Delta_{\pm} \equiv -G_{\pm}\sum_{\mathbf{k}}ke^{\pm i\phi_{\mathbf{k}}}\langle s_{\mathbf{k}}^{-}\rangle.
  \label{eq:def_delta_tr_broken}
\end{equation}
For $G_+=G_-$ this Hamiltonian turns into the full $p$-wave Hamiltonian Eq.~\eqref{eq:pwave_ham} we studied above, see Eq.~\eqref{eq:kdotq}.

 We consider quenches from a perturbed $p+ip$ initial state for the two cases
where one channel dominates over the other employing the same approach to dynamics as before.
For $G_{+} > G_{-}$, the dominant channel corresponds to the $p+ip$ channel.
Using numerical simulations we find that for any quench, the $p-ip$ order
parameter, $\Delta_{-}(t)$, quickly vanishes, as shown in
Fig.~\ref{fig:dom_right}. This result is to be expected as the $p-ip$ order
is suppressed in both the initial state as well as by the equations of motion
themselves.

For $G_{+} < G_{-}$, the dominant channel corresponds to the $p-ip$ channel and
it is not immediately clear which channel the dynamics should favor. On one
hand, the initial state favors the $p+ip$ channel while the $p-ip$ channel only
acts as a perturbation, on the other hand, the equations of motion suppress the
$p+ip$ channel and favor the $p-ip$ channel. We find that, in this case, the
$p+ip$ order parameter is the one to rapidly decay to zero while the $p-ip$
order parameter survives to exhibit the late time dynamics (in phase I both
order parameters decay to zero).
This behavior is shown in Fig.~\ref{fig:dom_wrong}, where even though
$\Delta_{-}(0)/\Delta_{+}(0) \ll 1$, the $p-ip$ order parameter amplitude is
the one with nonvanishing dynamics in phases II and III.
These late time dynamics resemble the three phases of the pure $p\pm ip$
quench dynamics but with a renormalized set of quench coordinates. It is an
interesting question as to whether the phase boundaries of the asymetric model,
particularly for $G_+ < G_-$, remain the same as in the chiral model or are deformed
in some way.

We see that the stronger channel always wins at late times regardless of the
initial state. This means that the time reversal invariant $p$-wave
Hamiltonian is an unstable fixed point with regards to the $p\pm ip$ phase
diagram. Only at the special point $G_{+} = G_{-}$ protected by time reversal
symmetry does the chaotic phase III' regime emerge, see
Fig.~\ref{fig:fixedpoint_flow}.

\section{Conclusion \label{sec:conclusion}}
In this work we have studied the quench dynamics of a $p$-wave BCS superfluid
with two competing order parameters, $\Delta_\pm(t)$.
We have shown that when the system is prepared \emph{near} its $p+ip$ ground state
and the interaction strength is quenched, the late time dynamics can be characterized
into three distinct phases:
in phase I both order parameters decay to zero,
in phase II $\Delta_+(t)$ decays to a nonzero constant and $\Delta_-(t)$ oscillates near zero,
and in phase III' the two order parameters are nonzero and undergo chaotic dynamics, see Fig.~\ref{fig:full_quenches}.
Remarkably, even though this model is nonintegrable, we are able to map out the exact phase boundaries
in parameter space as shown in Fig.~\ref{fig:full_PD}.
Additionally, we study the role that time reversal symmetry plays in determining the late time dynamics.
We consider a Hamiltonian that has an asymmetry in the coupling strength and prefers one of the order
parameters over the other, i.e., $G_+ \neq G_-$ in Eq.~\eqref{eq:bcs_asym_ham}.
We find that the late time dynamics of the weaker channel order parameter vanish and only the
stronger channel survives.
This causes the chaotic dynamics of phase III' to disappear and the topological phase III of the
chiral $p$-wave model studied in Ref.~\cite{Foster_2013} to emerge.
The fact that the stronger channel always wins suggests that it may be possible
to experimentally realize a quench induced Floquet topological superfluid
without any fine tuning of the initial state provided that
time reversal symmetry can be broken in the interaction channels.

Although our analysis has only been performed for the $p$-wave BCS superfluid, we expect a similar chaotic phase to emerge in far from equilibrium dynamics of other models where there are competing order parameters related by a  symmetry, for example, in a Fermi gas with more than two species of fermions with pairwise attraction  between them \cite{HH,
Cherng2007SuperfluidityAM, PhysRevA.78.033615, PhysRevA.81.033629}.

\begin{acknowledgments}
We thank M. Dzero, M. S. Foster, and V. Gurarie for helpful discussions. A.\
Z.\ is partially supported by Grant No. 2018058 from the United States-Israel
Binational Science Foundation (BSF), Jerusalem, Israel and through a Fellowship
from the Rutgers Discovery Informatics Institute.
\end{acknowledgments}

\onecolumngrid
\appendix
\section{Details of numerical simulations \label{apdx:num_sim}}
The numerical simulations were done using the parameters listed in
Table~\ref{tbl:pwave_params} and Table~\ref{tbl:chiral_params}.
For the full $p$-wave simulations, we use a radial momentum space grid so that
$\epsilon = k^2$ are uniformly spaced for $\epsilon \in \left[0, 2\Lambda\right]$
and along the angular direction $\theta$ is uniformly spaced in the upper half plane for $\theta \in \left[0,\pi\right]$.
For the chiral dynamics, the problem reduces to one dimension and we can neglect the $\theta$ dependence.
The number of spins
used in the numerical simulations is chosen such that the results are converged
at the times of interest.
We find that to obtain converging results at late times, it is important to have a
large number of points along the radial direction, $N_\epsilon$,
whereas the results are not as sensitive to the number of points
along the angular direction, $N_\theta$.
Since we have few $N_\epsilon$ points in the full $p$-wave model, we use the composite Simpson's rule to have a
more accurate estimate of the integrals in the problem while for the chiral model we simply use the trapezoidal rule.
Obtaining convergent results becomes difficult for quenches in phase III'
where the dynamics are chaotic. In order to reach later times we must increase
$N_\epsilon$, but changing the number of points effectively changes the initial conditions.
The results, however, are qualitatively the same, i.e., the Floquet phase is destroyed
and irregular oscillations appear.

\begin{table}[h!]
  \begin{tabular}{l c c}
    \hline \hline
    Quantity                    & Symbol         & Value       \\
    \hline
    Number of $\epsilon$ points & $N_{\epsilon}$ & 8001        \\
    Number of $\theta$ points   & $N_{\theta}$   & 25          \\
    Density                     & $n$            & 0.825       \\
    Fermi Energy                     & $\epsilon_F$            & $2\pi n$       \\
    High energy cutoff          & $\Lambda$      & $50 \epsilon_F$ \\
    Quantum critical point          & $\Delta_\mathrm{QCP}$      & $\approx 1.536$\\
    Ground state order parameter amplitude          & $\Delta_{0,+}$      & See figures for details \\
    $\Delta_-$ perturbation          & $\Delta_{0,-}$      & $10^{-5}\Delta_{0,+}$ \\
    \hline \hline
  \end{tabular}
  \centering
  \caption{Parameters used to simulate the full $p$-wave dynamics described by Eq.~\eqref{eq:eom}}
  \label{tbl:pwave_params}
\end{table}
\begin{table}[h!]
  \begin{tabular}{l c c}
    \hline \hline
    Quantity                    & Symbol         & Value       \\
    \hline
    Number of $\epsilon$ points & $N_{\epsilon}$ & 50000        \\
    Density                     & $n$            & 0.825       \\
    Fermi Energy                     & $\epsilon_F$            & $2\pi n$       \\
    High energy cutoff          & $\Lambda$      & $50 \epsilon_F$ \\
    Quantum critical point          & $\Delta_\mathrm{QCP}$      & $\approx 1.536$\\
    Ground state order parameter amplitude          & $\Delta_{0,+}$      & See figures for details \\
    \hline \hline
  \end{tabular}
  \centering
  \caption{Parameters used to simulate the $p+ip$ dynamics described by Eq.~\eqref{eq:eom_chiral}}
  \label{tbl:chiral_params}
\end{table}

\section{BCS equations\label{apdx:integrals}}
The integral on the right hand side of Eq.~\eqref{eq:mf_N_cont} for the particle density can be evaluated exactly
\begin{equation}
  n = \frac{1}{8\pi}\left( 2\Lambda + \lvert 2\mu\rvert -\sqrt{(2\Lambda-2\mu)^{2} + 8\Delta_{+}^{2}\Lambda} +
  2\Delta_{+}^{2} \ln \left[ \frac{2\Lambda -2\mu + 2\Delta_{+}^{2} + \sqrt{(2\Lambda-2\mu)^{2} + 8\Delta_{+}^{2}\Lambda}}{2\Delta_{+}^{2} + \lvert 2\mu \rvert - 2\mu}\right]\right).
\end{equation}
Expanding the square root in $1/\Lambda$ and neglecting terms of order $\Lambda^{-1}$ and smaller leads to
\begin{equation}
  4\pi n =  2\mu\Theta(\mu) - \Delta^{2}_{+} +
  \Delta_{+}^{2}\ln\left[\frac{2\Lambda}{\Delta_{+}^{2} + 2\lvert\mu\rvert\Theta(-\mu)}\right],
  \label{eq:n_expanded}
\end{equation}
Similarly,  Eq.~\eqref{eq:mf_D_cont}  becomes
 \begin{equation}
  \frac{2\pi}{g} = 2\Lambda - 8\pi n +
    2\mu\ln\left[\frac{2\Lambda}{\Delta_{+}^{2} + 2\lvert\mu\rvert\Theta(-\mu)} \right].
\end{equation}
We use these results in Appendix~\ref{apdx:lax} to derive the phase boundaries.

\section{Lax roots and quench phase diagram\label{apdx:lax}}
In the $p+ip$ model, the components of the Lax vector are given by~\cite{Foster_2013}
\begin{align}
  L^{\pm}(u) = \sum_{i=1}^{N} \frac{\sqrt{\epsilon_{i}}s_{i}^{\pm}}{\epsilon_{i}-u}, \quad
  L^{z}(u) = \sum_{i=1}^{N} \frac{\epsilon_{i}s_{i}^{z}}{\epsilon_{i}-u} + \frac{1}{2G_{f}}
\end{align}
and one can define a Lax vector norm
\begin{equation}
  L_{2}(u) = u L^{+}(u)L^{-}(u) + [L^{z}(u)]^{2},
\end{equation}
that is conserved under the dynamics. The Lax vector norm is a polynomial of
degree $2N$ whose isolated roots encode information about the late time dynamics~\cite{yuzbashyan2006relaxation, yuzbashyan2006dynamical, Foster_2013,yuzbashyan2015quantum}.
Specifically, phase I occurs when $L_2(u)$ has  no isolated roots (the remaining roots form a continuum
on the positive real axis). Phase II corresponds to a single pair and phase III to two pairs of isolated roots.
To find the roots, we can evaluate $L_2(u)$ in the initial state which has the
configuration described by Eq.~\eqref{eq:spin_eq} with $\Delta_{0,-} = 0$ and $\Delta_{0,+} \equiv \Delta_i$, i.e.,
\begin{align}
  \begin{split}
  s_{i}^{-} = -\frac{\sqrt{\epsilon_{i}}\Delta_{i}}{E(\epsilon_{i})}&, \qquad
  s_{i}^{z} = -\frac{(\epsilon_{i}-2\mu_{i})}{2 E(\epsilon_{i})}, \\
  E(\epsilon) = &\sqrt{(\epsilon - 2\mu_{i})^{2} + 4\epsilon\lvert\Delta_{i}\rvert^{2}}.
  \end{split}
\end{align}
Substituting into the above equations, the Lax vector norm becomes
\begin{equation}
  L_{2}(u) = u\lvert\Delta_{i}\rvert^{2}\left[\sum_{i}\frac{\epsilon_{i}}{(\epsilon_{i}-u)E(\epsilon_{i})} \right]^{2} +
  \left[-\sum_{i}\frac{\epsilon_{i}(\epsilon_{i}-2\mu_{i})}{2(\epsilon_{i}-u)E(\epsilon_{i})} + \frac{1}{2G_{f}} \right]^{2}.
\end{equation}
This equation can be simplified to
\begin{equation}
  L_{2}(u) = u\lvert\Delta_{i}\rvert^{2}\left[F(u) \right]^{2} +
             \left[-\frac{(u - 2\mu_i)}{2}F(u) + \frac{\tilde{\beta}}{2} \right]^{2}
\end{equation}
by defining
\begin{equation}
  \tilde{\beta} \equiv \frac{1}{G_{f}} - \frac{1}{G_{i}},
  \quad F(u) \equiv \sum_{i}\frac{\epsilon_{i}}{(\epsilon_{i}-u)E(\epsilon_{i})}.
\end{equation}
The roots of the Lax vector norm are found by solving the quadratic equation for $F(u)$
\begin{equation}
           L_{2}(u) = E(u)^2 F(u)^2 -2(u - 2\mu_i)\tilde{\beta} F(u) + \tilde{\beta}^2 = 0,
\end{equation}
with solutions given by
\begin{equation}
  F(u) = \frac{(u-2\mu_{i})\tilde{\beta} \pm 2i\lvert\Delta_i\rvert\lvert\tilde{\beta}\rvert\sqrt{u}}{E(u)^{2}}.
  \label{eq:roots_discrete}
\end{equation}
Isolated roots are located away from the positive real axis. Eq.~\eqref{eq:roots_discrete} for such roots in the continuum limit becomes
\begin{equation}
    \int_0^{2\Lambda} \frac{\epsilon\mathrm{d}\epsilon}{(\epsilon-u)E(\epsilon)} =
    \frac{(u-2\mu_{i})\beta \pm 2i\lvert\Delta_{i}\rvert\lvert\beta\rvert\sqrt{u}}{E(u)^{2}},
    \label{eq:cont_roots}
\end{equation}
with
\begin{equation}
\beta = 2\pi\left(\frac{1}{g_{f}} -\frac{1}{g_{i}}\right).
\label{eq:beta}
\end{equation}
To determine the phase boundaries we look for a pair of complex roots that just
separate from (or collapse to) the real axis. To do this, we replace
$u\rightarrow u \pm i\delta$ in Eq.~\eqref{eq:cont_roots}, with $\delta$
infinitesimally small. Through a change of variables $x = \epsilon - u$ we may
write
\begin{equation}
  \int_{-u}^{2\Lambda-u} \frac{(x+u)\mathrm{d}x}{(x \mp i\delta)E(x+u)} =
  \mathcal{P}\int \frac{(x+u)\mathrm{d}x}{xE(x+u)} \pm i\pi\frac{u}{E(u)}
  \label{eq:sok_plem}
\end{equation}
where $\mathcal{P}$ denotes the principal value.

Comparing the imaginary parts of the right hand sides of
Eq.~\eqref{eq:sok_plem} and Eq.~\eqref{eq:cont_roots} we find
\begin{equation}
\lvert\beta\rvert = \frac{\pi \sqrt{u}E(u)}{2\lvert\Delta_i\rvert},
\label{eq:root1}
\end{equation}
and comparing the real parts we find
\begin{equation}
  \ln\left[\frac{2\Lambda}{\Delta_i^2 +2\lvert\mu_i\rvert\Theta(-\mu_i)}\right] +
  \frac{u}{E(u)} \ln\left[-\frac{2\left[u(\Delta_i^2-\mu_i)+2\mu_i^2 + \lvert \mu_{i}\rvert E(u)\right]}
    {u\left[u + 2\Delta_i^2 - 2\mu_{i} + E(u)\right]}\right] =
  \mathrm{sgn}(\beta)\frac{\pi\sqrt{u}(u - 2\mu_i)}{2\lvert\Delta_i\rvert E(u)},
\label{eq:root2}
\end{equation}
where we have neglected terms of order $O(\Lambda^{-1})$.

Equations~\eqref{eq:beta},~\eqref{eq:root1}, and~\eqref{eq:root2} form a system
of equations that paramaterize the phase boundaries. By choosing a value for
$\Delta_{i}$ and $\mathrm{sgn}(\beta)$, Eq.~\eqref{eq:root2} can be solved for
$u$ which can be substituted into Eq.~\eqref{eq:root1} to determine $\beta$.
Then, with the help of Eq.~\eqref{eq:beta}, $\Delta_{i}$ can be written as a function of
$\Delta_{f}$ to map out the phase boundaries. This is the approach we used to obtain the phase boundaries shown in Fig.~\ref{fig:full_PD}.

\section{Nonequilibrium extension of the quantum critical point\label{apdx:beta_qcp}}
The nonequilibrium extension of the quantum critical point corresponds to the
curve in phase II which has $\mu_{\infty} = 0$. This line separates the two
topological regions with field winding number, $W$, either zero or one. To
determine the curve we look for the vanishing of an isolated root i.e., $u=0$~\cite{Foster_2013}.
Substituting $u=0$ into Eq.~\eqref{eq:cont_roots} we have
\begin{equation}
  \int_{0}^{2\Lambda}\frac{\epsilon \mathrm{d}\epsilon}{\epsilon E(\epsilon)} = -\frac{\beta}{2\mu_{i}}.
\end{equation}
Evaluating the integral we find
\begin{equation}
  \beta = -2\mu_{i}\log\left[\frac{2\Lambda}{\Delta_{i}^{2} + 2\lvert\mu_{i}\rvert\Theta(-\mu_{i})}\right].
\end{equation}
As before, we can use Eq.~\eqref{eq:beta} to write $\Delta_{i}$ as a function of
$\Delta_{f}$. The result simplifies to
\begin{equation}
  \mu_f\log\left[\frac{2\Lambda}{\Delta_f^2 + 2\lvert\mu_f\rvert\Theta(-\mu_f)}\right] = 0
\end{equation}
which can only be solved when $\mu_f = 0$. In other words, for any $\Delta_i$
the value of $\Delta_f$ for which $\mu_\infty = 0$ is given by
$\Delta_f = \Delta_\mathrm{QCP}$, see the dashed line in the phase diagram in Fig.~\ref{fig:full_PD}.
This line remains unchanged in the full $p$-wave phase diagram since it lies entirely within the stable phase II.

\twocolumngrid
\bibliography{references}
\end{document}